\documentclass[aps,prl,twocolumn,superscriptaddress,fleqn,a4paper]{revtex4-1}
\usepackage[ngerman,english]{babel}
\usepackage[utf8]{inputenc}   
\usepackage{xspace}
\usepackage{amsmath,tensor}
\usepackage{array,color}
\usepackage{relsize,exscale}  
\usepackage{textcase}	
\usepackage{soul,color}		
\usepackage{calc} 
\usepackage{makeidx,showidx}
\usepackage{graphicx}
\usepackage{wrapfig}
\usepackage{float}
\usepackage{textcomp}
\usepackage{siunitx}
\usepackage[version=3]{mhchem}
\usepackage{enumitem}
\usepackage{fancyref}
\usepackage[hidelinks=true]{hyperref}
\sethlcolor{green}
\setulcolor{red}
\usepackage{lineno}			

\newcommand{\erww} [1] {\ensuremath{\langle {#1} \rangle}}

\newcommand{\lsco} {{La$_{2-x}$Sr$_x$CuO$_4$}\@\xspace}

\newcommand{\hgryb} {{HgBa$_{2}$CuO$_{4+\delta}$}\@\xspace}

\newcommand{\ybco} {$\ce{YBa2Cu3O_{6+y}}$\@\xspace}

\newcommand{\ybcoE} {$\ce{YBa2Cu4O8}$\@\xspace}

\newcommand{\tc} {\ensuremath{T_{\rm c}}\@\xspace}
\newcommand{\cperp}{\ensuremath{c \bot B_0}\@\xspace}
\newcommand{\cpara}{\ensuremath{{c\parallel\xspace B_0}}\@\xspace}

\newcounter{exex}[section]

\makeatletter\newcommand\listofexamples{\section*{List of Examples}\@starttoc{xmp}}
\newcommand\l@example[2]{\par\noindent#1~\textit{#2}\par}
\makeatother






\makeatletter
\renewcommand\subsection{\@startsection 
	{subsection}{3}{0mm}
	{-\baselineskip}
	{0.5\baselineskip}
	{\centering \textbf }}
\makeatother

\makeatletter
\renewcommand\subsubsection{\@startsection 
	{subsubsection}{3}{0mm}
	{-\baselineskip}
	{0.5\baselineskip}
	{\centering  }}
\makeatother

\makeindex

\begin{document}




\title{Planar Cu and O NMR and the Pseudogap of Cuprate Superconductors}

\date{\today}
\author{Marija Avramovska}
\author{Jakob Nachtigal}
\author{Stefan Tsankov}
\author{J\"urgen Haase}


\affiliation{%
Felix Bloch Institute for Solid State Physics, University of Leipzig,  Linn\'estr. 5, 04103 Leipzig, Germany\\
}





\begin{abstract}
Recently, an analysis of all available planar oxygen shift and relaxation data for the cuprate high-temperature superconductors showed that the data can be understood with a simple spin susceptibility from a metallic density of states common to all cuprates. It carries a doping dependent but temperature independent pseudogap at the Fermi surface, which causes the deviations from normal metallic behavior, also in the specific heat. Here, a more coherent, unbiased assessment of all data, including planar Cu, is presented and consequences are discussed, since the planar Cu data were collected and analyzed prior to the O data. The main finding is that the planar Cu shifts for one direction of the external magnetic field largely follow from the same states and pseudogap. This explains the shift suppression stated more recently, which leads to the failure of the Korringa relation in contrast to an enhancement of the relaxation due to antiferromagnetic spin fluctuations originally proposed. However, there is still the need for a second spin component that appears to be associated with the Cu $3d(x^2-y^2)$ hole to explain the complex Cu shift anisotropy and family dependence.  Furthermore, it is argued that the planar Cu relaxation which was reported recently to be rather ubiquitous for the cuprates, must be related to this universal density of states and the second spin component, while not being affected by the simple pseudogap. Thus, while this universal metallic density of states with a pseudogap is also found in the planar Cu data, there is still need for a more elaborate scenario that eludes planar O.
\end{abstract}


\maketitle


\section{Introduction}
Nuclear magnetic resonance (NMR) played an important role in high-temperature superconductivity \cite{Bednorz1986}, in particular in its early days with the focus on the \lsco and \ybco families of materials. Important information about chemical and electronic properties could be obtained, but an unchallenged interpretation was not achieved, as of today, due to conflicting experimental evidence (for an early review see \cite{Slichter2007}). In particular the question of the description of the cuprates in terms of a single spin component electronic susceptibility remained unanswered, as more recent experiments had shown \cite{Haase2009b,Haase2012,Rybicki2015}.

Two very recent publications \cite{Nachtigal2020,Avramovska2021} revealed that \emph{all} planar oxygen NMR relaxation and shift data available from the literature for hole-doped cuprates (more than 60 independent data sets in total) are in agreement with a simple metallic spin susceptibility from a density of states that has a temperature independent pseudogap at the Fermi surface. The pseudogap opens as doping decreases from high levels and can be measured with NMR. The density of states outside the gap or in its absence is \emph{universal} to the cuprates, independent of doping and family \cite{Nachtigal2020}. This scenario is similar to what was concluded from specific heat data \cite{Loram1998,Tallon2020}. Thus, lowering the doping opens the gap, and the lost, low-energy states cease to contribute to shift, relaxation, or specific heat. As a consequence, one observes the following in NMR: (i) a high-temperature, doping dependent offset in the relaxation rate, $1/T_1$, that remains metal like in the sense that $\Delta(1/T_1)/\Delta T = const.$, irrespective of material and doping, and (ii) a high-temperature, doping dependent offset in the spin shifts, since electronic polarization from the low energy states is still missing even at the highest $T$ in the presence of the pseudogap. The states lost near the Fermi surface, in connection with thermal excitations across the gap, cause the typical temperature dependence of the spin shift hitherto ascribed to the opening of the gap (spin gap) at a given temperature. While the gap develops with doping as described, there are family differences, as well. For example, for \ybco the gap is almost closed at optimal doping, while for \lsco the gap is still sizable at optimal doping. In fact, if one uses $\zeta$, the doping measured with NMR from the charges at planar Cu and O, with $\zeta = n_\mathrm{Cu} + 2n_\mathrm{O}-1$ \cite{Jurkutat2014}, this is not surprising as optimally doped \ybco is found to have $\zeta \approx 20\%$, significantly larger than what is typically assumed. Again, this family dependence does not concern the density of states.

The question arises, how planar Cu NMR data relate to this picture. In fact, some of us began collecting and reanalyzing the planar Cu data prior to the planar O data, and accounts have been published \cite{Haase2017,Avramovska2019,Jurkutat2019}. For example, it was found that above \tc there must be a \emph{suppression} of the shifts rather than an enhancement of the nuclear relaxation \cite{Avramovska2019} since the Cu shifts vary widely while the Cu relaxation does not (in terms of $1/T_1T$) \cite{Jurkutat2019}. This is contrary to what was believed to be the reason for the failure of the Korringa relation between shift and relaxation \cite{Slichter2007}. With the new planar O analysis, we recognize immediately that the shift suppression is due to the pseudogap, but it does not affect the Cu relaxation.
Here we set out to develop a more coherent assessment of all new findings.
First, we discuss the overall phenomenology of the magnetic shifts of planar Cu and O. Then, we take a closer look at the assumptions about the hyperfine scenarios and orbital shifts, before we discuss consequences. Thereafter, we turn to the relaxation data, and their explanation in the new scenario.\par\medskip
\section{Planar C\lowercase{u} and O Magnetic Shifts}
\subsection{Overview}\label{sec:over}
\begin{figure*}[t]\centering
	\includegraphics[width=0.9\textwidth ]{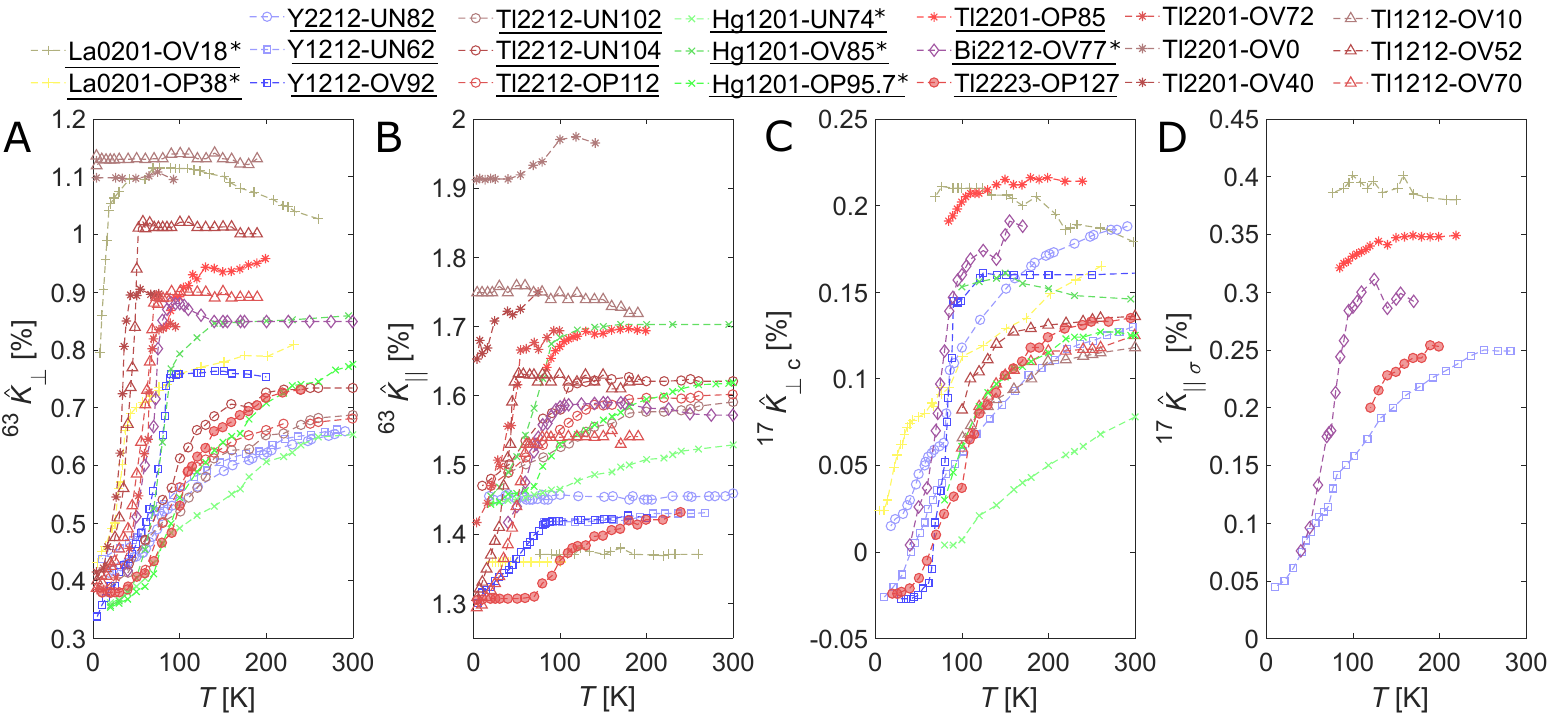}
	\caption{Total magnetic shifts for planar Cu and O as a function of temperature, (A) ${^{63}\hat{K}}_\perp(T)$, (B) ${^{63}\hat{K}}_\parallel(T)$, (C) ${^{17}\hat{K}}_{\perp\mathrm{c}}(T)$, and (D) ${^{17}\hat{K}}_{\parallel\sigma}(T)$. The vertical axis was adjusted in offset and scale so that the temperature dependences fill about the same space. These are representative data sets of the recently given full accounts \cite{Haase2017,Nachtigal2020}. The star (*) denotes samples where $^{63}$Cu and $^{17}$O data had to be taken from different publications and samples (slightly different $T_\mathrm{c}$ were reported);  an underlined legend entry highlights a material (from \cite{Haase2017}) for which oxygen data are available. Note that ${^{63}\hat{K}}_{\parallel,\perp}$ denotes the planar Cu total magnetic shifts measured with the magnetic field parallel and perpendicular to the crystal $c$-axis (the in-plane direction is often not known); for planar O the total magnetic shifts are specified for the field direction with respect to the $\sigma$-bonding between Cu and O: for ${^{17}\hat{K}}_{\parallel\sigma}$ the field is along the sigma bond, while for ${^{17}\hat{K}}_{\perp c}$ the field is perpendicular to the $\sigma$-bond and parallel to the crystal $c$ axis; with the field along the $\sigma$-axis ${^{17}\hat{K}}_{\perp a}$ can be measured as a 2nd planar O peak.}\label{fig:fig1}
\end{figure*}

We begin with an overview of the experimental shift data, and in order to keep the discussion transparent for the reader we present the bare magnetic shifts that still include, in particular, the van Vleck orbital contribution (the shielding from the core is absent due to shift referencing, for more details see \cite{Haase2017}). The data are presented in Fig.~\ref{fig:fig1} and one can draw some
apparent, fundamental conclusions: (1) All Cu and O shifts show similar temperature dependences. (2) The range of the temperature dependence for the Cu shifts (panels A and B) is similar for both directions of the magnetic field, about 0.75\%. (3) For planar O, the two ranges are different, 0.24\% and 0.35\% perpendicular and parallel to the $\sigma$-bond, respectively (panels C and D). (4) The low temperature shifts for planar Cu are large, 0.3\% and 1.3\% (supposed to be the orbital shifts), while they play a minor role for planar O.

Based on these observations one is inclined to conclude that one uniform spin susceptibility rules the temperature dependent shifts for both nuclei, and that planar Cu is coupled to the related electronic spin by an \emph{isotropic} hyperfine coupling constant while that for planar O has an anisotropy of about 1.45. 
Given a partially filled Cu $3d(x^2-y^2)$ orbital that is $\sigma$-bonded with four O $2p$, the above findings are expected for O, but not at all for Cu, for two reasons. First, one expects a large anisotropy for the hyperfine coefficient for spin in the $3d(x^2-y^2)$ orbital (a factor of at least 6 \cite{Pennington1989}, see also below). Second, the orbital shifts, that follow from this in the low temperature limit, and their anisotropy do not fit this bonding scenario. The hybridization with planar O should result in an orbital shift anisotropy significantly less than 4, the single ion value \cite{Pennington1989} (see also below). Finally, a closer look at Fig.~\ref{fig:fig1}{\bf B} reveals that the planar Cu shifts measured with the $B_0$ field parallel to the crystal's $c$-axis (\cpara) show a distinct family dependence that violates the overall view. Some materials seem to have a vanishing shift range for \cpara (\lsco) while others occupy a different part of the panel (this family dependence is similar to what was found based on the charge sharing in the CuO$_2$ plane \cite{Jurkutat2014}).

Before we continue with a more detailed discussion, we remind the reader of the single spin component picture.

\subsection{Single Spin Component Picture}
In case of a simple electronic spin susceptibility, $\chi(T)$,  that leads to a uniform electronic spin component, $\erww{S_z}$, in linear response to the applied external magnetic field, $B_0$, we have, 
\begin{equation}\label{eq:spin}
\erww{S_z}(T) = \chi(T) \;B_0/\gamma_e \hbar.
\end{equation}
And all (non-trivial) spin shifts, ${^{n}K}_{\theta}$, observed at a nucleus, $n$, for different orientations, ${\theta}$, of the external magnetic field with respect to the crystal axes must then be proportional to $\erww{S_z}$, and thus to each other. The proportionally constants define the hyperfine coefficients, ${^{n}H}_{\theta}$. Note, however, with NMR we measure the total magnetic shift (as given in Fig.~\ref{fig:fig1}),
\begin{equation}\label{eq:shift}
{^{n}\hat{K}}_{{\theta}}(T) =  {^{n}K}_{{\rm L}{\theta}} + {^{n}H}_{\theta} \cdot \chi(T),
\end{equation}
where the orbital (chemical) shift term, $^{n}K_{{\rm L}{\theta}}$, follows from the orbital (van Vleck) susceptibility that is expected to be temperature independent (but is very different for different nuclei, reflecting the bonding). The spin shift is given by,
\begin{equation}\label{eq:spinsingle}
{^{n}K}_{{\theta}}(T) = {^{n}H}_{\theta} \cdot \chi(T).
\end{equation}
Then, if we plot one shift vs.\@ another, for every measured temperature point, $T_i$, (i.e. ${^{n}\hat{K}}_{\theta_1}(T_i)$ vs.~${^{k}\hat{K}}_{\theta_2}(T_i)$), we must observe straight lines with slope ${^{k}H}_{\theta_1}/{^{n}H}_{\theta_2}$, with an offset given by the orbital shifts, and with \eqref{eq:shift} and \eqref{eq:spinsingle}, we get,
\begin{equation}\label{eq:single}
{^{n}\hat{K}}_{{\theta_1}}(T) =  \frac{{^{n}H}_{\theta_1}}{{^{k}H}_{\theta_2}} \cdot {^{k}\hat{K}}_{\theta_2}(T) + \left[{^{n}K}_{L\theta_1}-\frac{{^{n}H}_{\theta_1}}{{^{k}H}_{\theta_2}} \cdot{^{k}K}_{L\theta_2}\right].
\end{equation}

\subsection{Hyperfine Coefficients and Orbital Shifts}\label{sec:hyper}
Before we discuss the shifts in more detail, we revisit the assumptions about the hyperfine scenarios and orbital shifts. 

First, we address the $^{17}$O NMR. Here, one expects from the bonding in the CuO$_2$ plane a (traceless) dipolar hyperfine tensor, with principle values $c_{\mathrm{dip},\theta}$, and an isotropic term $c_\mathrm{iso}$, from the $2p$ and $2s$ orbitals, respectively. The dipolar tensor will have its largest principle  value along the $\sigma$-bond and it is expected to be symmetric with respect to the other two main axes, and one has for the spin shift,
\begin{equation}\label{eq:ox1}
{^{17}K}_{\theta}(T) = \left[c_\mathrm{iso} + c_{\mathrm{dip},\theta}\right] \cdot \chi (T) \equiv C_\theta \cdot \chi (T).
\end{equation}
The orbital shifts are expected to be small so that \eqref{eq:shift} should be well approximated by \eqref{eq:ox1}. This is indeed the case \cite{Avramovska2021} and by applying \eqref{eq:single} one finds small orbital shifts that coincide with values predicted by first principles \cite{Renold2003}. From the slopes one infers hyperfine coefficients that are again in agreement with first principle calculations \cite{Huesser2000,Avramovska2021}. 

\par\medskip 

For planar Cu the situation is more complicated. Early on, one expected a large but negative  ${^{63}K}_\parallel$ from spin of the partly filled 3$d(x^2-y^2)$ orbital, as well as a large anisotropy due to the related hyperfine coefficient,  $|A_\perp| \lesssim  |A_\parallel|/6$, the sign of $A_\perp$ is not known with certainty \cite{Pennington1989,Huesser2000}. However, the shifts in Fig.~\ref{fig:fig1} are positive. So it was suggested that a transfer of spin density from the 4 neighboring Cu ions (in a single band picture) could be the reason for a positive hyperfine interaction \cite{Mila1989b}, i.e.,
\begin{equation}
{^{63}H}_{\parallel,\perp} = A_{\parallel,\perp} + 4 B,
\end{equation}
where $B$ is the related (positive) isotropic constant from transferred spin. With a single spin component one concluded on,
\begin{equation}\label{eq:singlecu}
{^{63}K}_{\parallel,\perp}(T) = \left[A_{\parallel,\perp} + 4 B\right] \cdot \chi (T).
\end{equation}
Then, since for \lsco and \ybco the shifts for \cpara are temperature independent, the famous accidental cancellation was invoked,
\begin{equation}\label{eq:cancel}
A_\parallel + 4 B \approx 0.
\end{equation}
In view of Fig.~\ref{fig:fig1}{\bf B} this is clearly not applicable for all cuprates, as ${^{63}K}_{\parallel}(T)$ can be as large as ${^{63}K}_{\perp}(T)$. The assumption \eqref{eq:cancel} has also consequences for the nuclear relaxation from field fluctuations parallel to the crystal $c$-axis: only fluctuations near the antiferromagnetic wave vector can contribute. With other words, any relaxation due to fields along the $c$-axis are interpreted as fluctuations at the antiferromagnetic wave vector even if this is not the case.

If one reflects on this scenario with the data from Fig.~\ref{fig:fig1}, one may be inclined to conclude that the hyperfine constant $A_\parallel$ must be family dependent, such that $A_\parallel \approx -4B$ for \lsco, a slightly more positive value for \ybco, and $A_\parallel \approx 0$ for other families. These would be very large changes of $A_\parallel$ and we dismiss this explanation. Also, one would expect that $B$ should be affected, as well, if $A_\parallel$ changes significantly, but Fig.~\ref{fig:fig1} does not support this, neither larger changes of the planar O hyperfine term or Cu nuclear relaxation (see below).

The various expected contributions to the magnetic hyperfine constants were pointed out early on \cite{Pennington1989,Huesser2000}. Spin in the unfilled 3$d(x^2-y^2)$ shell will interact with the nucleus through a (anisotropic) spin-orbit coupling, $a_{\mathrm{so},\theta}$, a (traceless) dipolar term, $a_{\mathrm{dip},\theta}$, and a (isotropic) core polarization of the $s$-shells, $a_\mathrm{cp}$. These various contributions add up to the traditional 'core polarization' term, $A_\theta = a_{\mathrm{so},\theta} + a_{\mathrm{dip},\theta}+a_\mathrm{cp}$, which is known to be negative for \cpara and $|A_\parallel |/|A_\perp|  \gtrsim 6 $ for Cu$^{2+}$ ions \cite{Pennington1989}. One would also expect that $A_{\parallel,\perp}$ is independent on the materials.

First principle cluster calculations of the hyperfine interaction in \lsco \cite{Huesser2000} support the above view. $A_\parallel$ is found to be negative and its magnitude is about 6 times larger than that of $A_\perp$. While $|A_\perp|$ is small, its sign may not be certain (the spin-orbit interaction is not well known and does not follow from the calculations). A large isotropic and positive transferred term $B$ (that depends on the number of Cu atoms in the cluster calculations) is found to have, indeed, a similar size but an opposite sign compared to $A_\parallel$. The following numbers were given (in atomic units) \cite{Huesser2000}: $a_{\mathrm{so},\parallel} \approx +2.405, a_{\mathrm{so},\perp} \approx +0.427, a_{\mathrm{dip},\parallel} \approx -3.644, a_{\mathrm{dip},\perp}  \approx +1.82, a_{\mathrm{cp}}\approx -1.78, 4b \approx +2.86 $, thus in our nomenclature $A_\parallel \approx -3.02, A_\perp \approx +0.47$, $4B = +2.86$ so that $A_\parallel + 4B \approx 0$ (${{^{63}H}_\parallel} \approx -0.16$, ${{^{63}H}_\perp} \approx 3.33$). 

Another early assumption concerns the orbital shifts for planar Cu. In view of Fig.~\ref{fig:fig1}{\bf A} and {\bf B} it was assumed that it is given by the low-temperature shifts (since spin singlet pairing leads to the complete loss of spin shift). As Fig.~\ref{fig:fig1} reveals, however, there is only a common ${^{63}K}_{\mathrm{L}\perp} \approx 0.3\%$ while the low temperature values vary between families for \cpara. While this cannot be excluded, given the family dependent charge sharing between Cu and O \cite{Jurkutat2014}, the fact that it does not affect ${^{63}K}_{\mathrm{L}\perp}$ is surprising since the orbital shift anisotropy is expected to be given by matrix elements and the overall symmetry, and both orbital shifts should be affected. In fact, first principle calculations for \lsco do predict the observed orbital shift for \cperp, ${^{63}K}_{\mathrm{L}\perp} \approx 0.30\%$. However, for \cpara the same calculations predict ${^{63}K}_{\mathrm{L}\parallel} \approx 0.72\%$, which is much smaller than what is observed even for \lsco, cf.~Fig.~\ref{fig:fig1}, where ${^{63}K}_{\mathrm{L}\parallel} \gtrsim 1.3\%$. Also on general grounds, since the single ion orbital shift has an anisotropy of 4, the hybridization with the planar O $\sigma$-bond should lead to a much smaller anisotropy (in support of the first-principle calculations \cite{Renold2003}).

\subsection{Two Spin Components}
Before we enter a deeper discussion, we give a few definitions for the uniform response in a two-component picture.

If a single spin susceptibility as in \eqref{eq:single} fails to explain the data, it may be useful to introduce another susceptibility. Thus, instead of \eqref{eq:spin}, we have two spin polarizations in the magnetic field, which can have different temperature and doping dependences,
\begin{align}
&\erww{S_\mathrm{1,z}}(T)= \chi_\mathrm{1}(T) \;B_0/\gamma_e \hbar\label{eq:twospin1}\\
& \erww{S_\mathrm{2,z}}(T) = \chi_\mathrm{2}(T) \;B_0/\gamma_e \hbar.\label{eq:twospin2}
\end{align}
 For the spin shifts one has instead of \eqref{eq:shift},
\begin{equation}\label{eq:shift2}
{^{n}\hat{K}}_\alpha(T)={^{n}K}_{\mathrm{L}\theta} + {^{n}H_1}_\theta \cdot \chi_1(T)+ {^{n}H_2}_\theta \cdot \chi_2(T),
\end{equation}
 since a nuclear spin couples to the two spin components with different hyperfine constants, ${^{n}H_{j\theta}}$, (if the two hyperfine constants were very similar, one would arrive at an effective single component description in \eqref{eq:shift2}). In general, the two spin components in \eqref{eq:twospin1} and \eqref{eq:twospin2} will have some sort of interaction, so that,
\begin{align}
\chi_1 &= \chi_{11} + \chi_{12} \label{eq:sus21}\\
\chi_2 &= \chi_{22} + \chi_{12},\label{eq:sus22}
\end{align}
(note that an antiferromagnetic coupling $\chi_{12}$ between the two spins with $\chi_{11}$ and $\chi_{22}$ could reduce the overall uniform spin response, $\chi_1 + \chi_2 = \chi_{11} + \chi_{22} + 2 \chi_{12}$, significantly). We will simplify the nomenclature with the definitions: $a \equiv \chi_{11}, \;b \equiv \chi_{22},\; c\equiv \chi_{12}$ (as used before \cite{Avramovska2019}). Then we have with \eqref{eq:shift2} the following formal expression for the (total) magnetic shift observed at a nucleus ($n$),
\begin{equation}\label{eq:shift21}
{^{n}\hat{K}}_\theta(T)={^{n}K}_{\mathrm{L}\theta} + {^{n}H_1}_\theta \cdot \left[a(T)+c(T)\right]+ {^{n}H_2}_\theta \cdot 
\left[b(T)+c(T)\right].
\end{equation}

\subsection{Discussion of C\lowercase{u} and O Shifts}
\begin{figure*}\centering
	\includegraphics[width=0.9\textwidth]{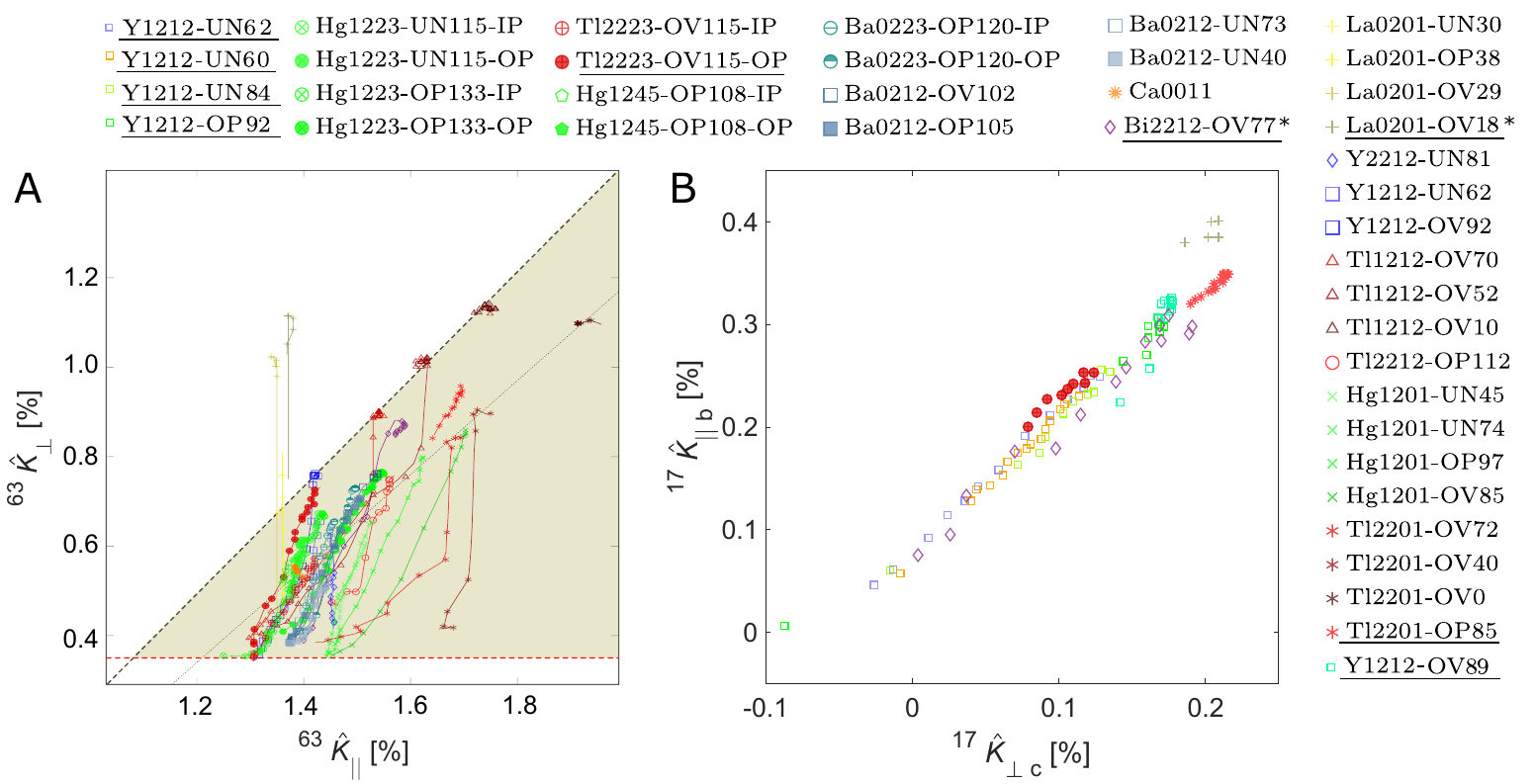}
	\caption{The complex shift anisotropy of planar Cu (A) is contrasted with the simple behavior found for planar O (B). The plot in (A) is adapted from Ref.\@ \cite{Haase2017} that also discusses the salient features. For example, 3 different segments of slopes seem to be necessary to explain the data. The steep slope is assumed by, e.g. \lsco but also other materials in certain ranges of temperature; the  "isotropic shift lines" (grey lines) have slope 1, i.e., both shifts change by the same amount as function of doping or temperature; finally, the slope 2.5 is assumed for a number of materials (only as a function of doping, not temperature). All  $^{63}$Cu shifts for $c\perp B_0$  have similar low $T$ shift values not far from $0.30\%$ calculated by first principles \cite{Renold2003}, however, the same calculations predict $0.72\%$ for \cpara, very different from what the experiments seem to show in (A). The planar O anisotropy is simple and the deduced small orbital shifts have been calculated, as well \cite{Renold2003}, and a simple anisotropic hyperfine coefficient explains the data (it is also in very good agreement with first principle calculations \cite{Huesser2000}).}
	\label{fig:fig2}
\end{figure*}
The planar O data, at all doping levels, are in agreement with a single spin component from a metallic density of states with a temperature independent pseudogap at the Fermi surface. The experimentally determined orbital shifts and hyperfine coefficients are in accord with first principles calculations \cite{Nachtigal2020,Avramovska2021,Renold2003,Huesser2000}. Despite this convincing, simple picture, we should point out that there is scatter between different systems above as well as below \tc, cf.~Fig.~\ref{fig:fig2}{\bf B}  and the discussion in \cite{Avramovska2021}. As pointed out previously \cite{Nachtigal2020,Pavicevic2020}, a spatial distribution of the planar O hole content, $n_\mathrm{O}$, that is often observed as NMR quadrupolar linewidth may cause a spatial variation of the pseudogap. In such a case the temperature dependence of the shift can change significantly as smaller gaps dominate at lower temperature. This fact, and the notorious consequences of the $^{17}$O isotope exchange harbor not well understood errors. Nevertheless, there is no obvious family dependence in terms of the temperature dependence of the shift, while the size of the pseudogap determined from fitting the planar O relaxation does depend on the materials to some extent \cite{Nachtigal2020}.\par\medskip 

The situation is very different for the planar Cu shifts. In view of Fig.~\ref{fig:fig1} one is inclined to conclude, as remarked above, that the planar Cu shifts appear to be dominated by an \emph{isotropic} hyperfine coefficient and a spin susceptibility with the temperature independent pseudogap that reigns the planar O data. The ranges of the shifts for both directions of the field are similar, $\Delta {^{63}K}_{\parallel,\perp}(T) \sim 0.75\%$, despite spin in the $3d(x^2-y^2)$, which should lead to a large shift anisotropy that is not found at all. A closer look reveals that there is a peculiar family dependence for \cpara: \lsco has the smallest and temperature independent shift, \ybco ranges a bit higher and has a slight temperature dependence, while other materials fill in the larger shift range, also with large temperature dependences. This family dependence bears similarities to that of how the inherent hole is shared between Cu ($n_\mathrm{Cu}$) and O ($n_\mathrm{O}$), and how doped holes enter the CuO$_2$ plane ($\Delta n_\mathrm{Cu}/\Delta n_\mathrm{O}$) \cite{Jurkutat2014}. We argued above that this cannot be caused by changes in $A_\parallel$. Consequently, an interplay between (at least) two spin components has to be considered.

In fact, if we take both Cu shifts in Fig.~\ref{fig:fig1} and plot them against each other, cf.~Fig.~\ref{fig:fig2}{\bf A}, we do not observe a straight line as expected from \eqref{eq:single} and seen in Fig.~\ref{fig:fig2}{\bf B} for planar O \cite{Avramovska2021}. Rather, for Cu we find three different slopes as a function of temperature and/or doping, as was pointed out with the original shift collection \cite{Haase2017}, 
\begin{equation}
\begin{split}\label{eq:slopes}
&\kappa_1 \approx \gtrsim 10, \;\;\;\kappa_2 \approx 1, \;\;\;\kappa_3 \approx 2.5, \text{ with}\\
 &\kappa_n = \Delta_{x, T} ^{63}K_\perp(T) /\Delta_{x, T} ^{63}K_\parallel(T).
\end{split}
\end{equation}
Note that this behavior also argues for spin effects as cause for the shift anisotropy since the three different slopes would require 3 different ratios of hyperfine coefficients in certain ranges of temperature or doping. 

Finally, there is the orbital shift conundrum which suggests that the old hyperfine scenario fails, and some of us suggested this can be solved by a negative spin polarization of the $3d(x^2-y^2)$ orbital (due to exchange with a second, positive component likely residing on planar O) \cite{Haase2017,Avramovska2019}. In principle, orbital currents  \cite{Varma2006} could represent or contribute to this shift. Since we see no clear scenario for this, we discuss the missing shift in terms of spin shift for the time being. That is, we set out to analyze the spin shifts in terms of \eqref{eq:shift21}, 
\begin{align}
{^{63}K}_{\theta}(T)&= A_{\theta} \cdot[a(T)+c(T)] + 4B \cdot[b(T)+c(T)],\label{eq:shift31}\\
{^{17}{K}}_{\theta}(T)&= C_{\theta} \cdot [b(T)+c(T)].\label{eq:shift32}
\end{align}
While we can view the above equations as the defining equations for the hyperfine coefficients, information about them can be drawn from  Fig.~\ref{fig:fig1}. Since $^{63}K_\perp(T)$ is similar to ${^{17}{K}}_{\theta}(T)$, one concludes that $|A_\perp|$ must be small compared to $|A_\parallel|$ and $4B$ (the transferred hyperfine coefficient should be isotropic). Note that such $A_{\parallel,\perp}$ is expected from an unfilled $3d(x^2-y^2)$ orbital. Thus, the choice above is reasonable.

With all planar O shifts and ${^{63}K}_\perp$ disappearing at low temperatures, $(b+c)$ approaches zero at low temperature: $(b_0+c_0) =0$. Furthermore, ${^{63}K}_\parallel (T\rightarrow 0) = A_\parallel (a_0 +c_0) \sim 0.6\%$ for a few systems and perhaps a bit larger for some other systems, cf.~Fig.~\ref{fig:fig1}, while the high temperature value of this contribution is given by $A_\parallel (a + c) \sim 0.3\%$, as one infers from the intersection of the high-temperature slope $\kappa_2 \approx 1$ with the abscissa(cf. \ref{fig:fig2}, \textbf{A}), ${^{63}\hat{K}}_\perp \approx 0.3\%$. Thus, $(a+c)$ grows more negative as the temperature is lowered, quite different from $(b+c)$ that vanishes at low temperatures. Some of us argued before \cite{Avramovska2019,Avramovska2020} about plausible assumptions with regard to \eqref{eq:shift31} and \eqref{eq:shift32}: $\kappa_2 = 1$ would result from a variation of $b$ only, and a large slope ($\kappa_1$) results if $4B/(A_\parallel+4B)$ is large. A variation of $a$ only would result in a very small slope, observed for just one system at very low temperatures, cf.~Fig.~\ref{fig:fig2}{\bf A}. Finally, the slope of $\kappa_3 = 2.5$ must involve more than one component.

Since first principle calculations support that $|A_\parallel| \sim 4B$, we will adopt this to approximate the Cu shifts by \eqref{eq:shift31} and \eqref{eq:shift32},
\begin{align}
{^{63}K}_\parallel(T)&\approx 4B\cdot [b-a]\label{eq:cu10}\\
{^{63}K}_\perp(T)& \approx 4B\cdot [b+c],\label{eq:cu11}
\end{align}
and we note that since $c$ is missing in \eqref{eq:cu10} its variation could be at the root of the interesting behavior of the shift anisotropy in Fig.~\ref{fig:fig2}{\bf A}. 
Unfortunately, one has three variables to explain the shifts in a two-dimensional plot.
\begin{figure*}\centering
\includegraphics[width=0.9\textwidth ]{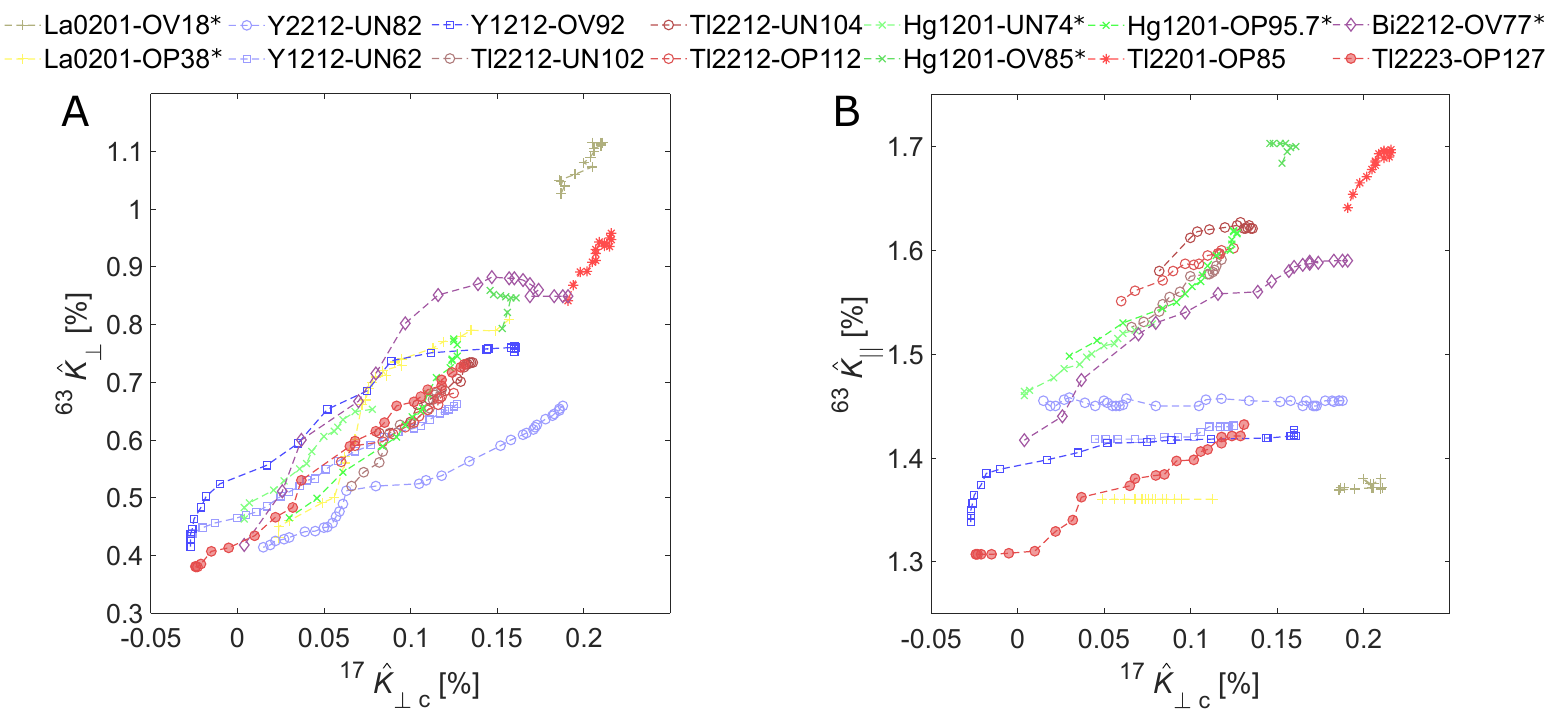}
\caption{$^{63}$Cu and $^{17}$O shift plotted against each other. (A), $^{63}\hat{K}_{\perp}$ vs. $^{17}\hat{K}_{\perp c}$, and (B),  $^{63}\hat{K}_{\parallel}$ vs. $^{17}\hat{K}_{\perp c}$. The differences between the two plots are in agreement with Fig.~\ref{fig:fig2}. Note that such plots were invoked for to prove the single fluid picture, early on, as well as its failure more recently (see main text for deeper discussion). The star (*) denotes samples where $^{63}$Cu and $^{17}$O data had to be taken from different publications and samples (slightly different $T_\mathrm{c}$ were reported)}
 \label{fig:fig3}
\end{figure*}

Finally, we notice that since ${^{63}K}_\perp(T)$ and ${^{17}K}_\theta(T)$ are functions of $[b+c]$ only, a plot of the two shifts with respect to each other should result in a straight line with a slope given by $4B/C \approx 3.0$, if one uses the the calculated hyperfine coefficients \cite{Huesser2000}. This plot is shown in Fig.~\ref{fig:fig3}{\bf A}, and holds a surprise. First of all, there is no well defined slope, rather, the temperature dependences appear somewhat erratic and lie between the two slopes of $4B/C \approx 2.0$ to 3.9. Note that the plot in panel {\bf B} does not hold new information if one considers the plot in Fig.~\ref{fig:fig2}{\bf A}, it is only shown for completeness. We must conclude that \eqref{eq:shift31} and \eqref{eq:shift32} only describe the situation on average. We cannot offer a clear explanation, but we would like to point to a few thoughts. In view of the behavior of planar O, cf.~Fig.~\ref{fig:fig2}{\bf B} (cf.\@ also discussion in \cite{Avramovska2021}), the uncertainty in the planar O shift cannot explain the observation. For planar Cu the linewidths can be very large, but even there we do not find a dependence that points to an uncertainty in the determination of the average shift (or its meaning). Note that $^{63}K_\perp$ follows after the subtraction of the material dependent quadrupole shift (in 2nd order) that may harbor unexpected errors, but the agreement with regard to the orbital shift, as well as the clear behavior in terms of only a few slopes in Fig.~\ref{fig:fig2}{\bf A} argue against this. Note that an overall change in spin density between O and Cu as a function of temperature could explain the finding, and we do know that the intra unit cell charger ordering seen with NMR \cite{Reichardt2018} can respond to temperature, but there is no evidence for a large change in the magnetic shift \cite{Reichardt2018}. However, such short range variations can lead to a different shift for Cu than for O (similar to the argument for charge ordering \cite{Haase2003}). Obvious differences between Cu and O arise from the two spin components. Apparently, planar O is not significantly affected by the Cu $3d(x^2-y^2)$ spin, only in terms of the description in terms of $(b+c)$, i.e. the coupling to it. Perhaps, the close proximity of both component on Cu do not follow such a simple description in detail (e.g. slight changes in the transferred coupling). 

It is important to remember, that such plots have been used, many years ago, to prove the so-called single fluid picture for YBa$_2$Cu$_3$O$_{6.63}$ \cite{Takigawa1991} and \ybcoE \cite{Bankay1994}. For both systems the dependences in Fig.~\ref{fig:fig3} may be taken as such, but in view of the dependences for other families, this conclusion cannot be taken as rigorous proof for all cuprates. Also, we now understand why similar plots for other systems, \lsco \cite{Haase2009b} and \hgryb \cite{Haase2012,Rybicki2015}, led to the opposite conclusion. We believe the latter still holds in view of the data, but the shift anisotropies in Fig.~\ref{fig:fig2} may be even more convincing.

\section{Planar C\lowercase{u} and O Nuclear Relaxation}
We pointed to the main facts about the Cu and O relaxation above, details have been published \cite{Avramovska2019,Jurkutat2019,Avramovska2020,Nachtigal2020}. The planar O relaxation of all cuprates follows from a density of states that is common to all cuprates and a doping dependent gap, the relaxation anisotropy is in agreement with the simple hyperfine scenario \eqref{eq:ox1}, as well. With the magnetic field along the $\sigma$-bond we have $1/T_{1\parallel\sigma}T \approx \SI{0.23}{/Ks}$ and this relates to a ${^{17}K}_{\perp c} \approx 0.2\%$, i.e. the Korringa relation holds as for a simple metal. 
For planar Cu the situation is quite different, yet also simple. The Cu relaxation rates \emph{do not} show pseudogap behavior at all. But the planar Cu relaxation rate just above \tc is common to all cuprates,  in the sense that $1/T_{1\perp} T_{\rm c} \approx \SI{21}{/Ks}$ (i.e. if measured with the field $B_0$ in the CuO$_2$ plane). \lsco is an outlier as its relaxation rate is about 3 times larger (\SI{60}{/Ks}). However, all cuprates have a temperature independent relaxation anisotropy (above and below \tc) and it varies between about 1.0 for some highly doped materials and about 3.4 for the \ybco family (\lsco has an anisotropy of about 2.4). Thus, the nuclear Cu spins are coupled by an anisotropic interaction to a metallic bath with a fixed density of states, as well (but no pseudogap). Importantly, below \tc all relaxation rates behave very similar in terms of $T/T_\mathrm{c}$,  as in the classical case (yet without a coherence peak) \cite{Jurkutat2019}. In fact, the planar Cu relaxation rate shows the sudden condensation at \tc very clearly, different from Cu shifts and planar O shift and relaxation since these are all dominated by the pseudogap.

A model for the planar Cu relaxation was discussed recently \cite{Avramovska2020} assuming two spin components $\alpha$ and $\beta$ that contribute to the fluctuating field at the Cu nucleus ($A_{\parallel,\perp}\cdot \alpha$ and $4B\cdot \beta$ with $A_\parallel =- 4B$ and $f = |A_\perp/A_\parallel|$). It was found that if both components are fully correlated, $\left<\alpha\cdot\beta\right>= +1$ and $\left<\beta\cdot\beta\right>= +1$, one can fit the planar Cu relaxation data and its anisotropy by varying $4\beta/\alpha$ from about $0.2$ to $0.4$. This means, a dominating spin component $\alpha$ that is correlated with the neighboring components $\beta$ can account for the observed relaxation and its anisotropy. Note that a normalized relaxation rate of $1/T_{1\perp}T=\SI{20}{/Ks}$ gives a planar Cu shift of  about 0.89\% if one invokes the Korringa relation. This is approximately the shift range of planar Cu.

Finally, we note that an antiferromagnetic coupling between two $\alpha$ components on adjacent Cu atoms frustrates the correlated spin $\beta$, unless they overcome the effective exchange coupling. This mechanism could be behind the observed pseudogap from the metallic density of states, and may not influence the planar Cu relaxation beyond a change in anisotropy.

\section{Conclusions}
To conclude, the recent collection of all available planar O shift and relaxation data \cite{Nachtigal2020} uncovered compelling arguments in favor of a simple metallic density of states with a temperature independent pseudogap set by doping. Even the anisotropies of O shift and relaxation \cite{Avramovska2021} support this simple view. The density of states outside the gap is common to all cuprates (independent on the size of the gap) and gives a metallic shift from the Korringa relation that is indeed observed. This raised the question of how this spin susceptibility relates to the phenomenology of the planar Cu data that had been published previously \cite{Haase2017,Avramovska2019,Jurkutat2019,Avramovska2020}. By comparing both sets of data we conclude that the planar Cu shifts are dominated by the same pseudogap, in particular, the earlier postulated suppression of the planar Cu shifts is a result of this pseudogap \cite{Avramovska2019}. However, there is a significant family dependence for the Cu shifts for one direction of the field (\cpara), so that this pseudogap feature can disappear for this direction of the field. As before, we endorse the conclusion that it is the action of a second spin component, likely from the spin in the $3d(x^2-y^2)$ Cu orbital, that is behind the complicated and family dependent planar Cu shift anisotropy. While the consequences of the pseudogap are obvious already in the bare Cu shifts, there is a quantitative difference between O and Cu that seems to be outside the expected uncertainties. It also carries a family dependence and explains why one can conclude on a single component picture for some systems, but not for other. Striking is still the observation of a doping and family independent metallic planar Cu relaxation that does not show a pseudogap at all \cite{Avramovska2019,Jurkutat2019}. One would argue that this must be related to the metallic density of states observed for planar O, perhaps as a result of antiferromagnetic coupling of the planar Cu spins, as argued before \cite{Avramovska2020}. 

 \section*{Acknowledgements}
 	We acknowledge the support by the German Science Foundation (DFG HA1893-18-1), and fruitful discussions by Anastasia Aristova, Robin Guehne, Andreas P\"oppl, Daniel Bandur. In particular we thank B. Fine (Leipzig) and A. Erb (Munich) for their extensive communications, and S. Kivelson (Stanford) for sharing ideas with JH.

 \section*{Author contribution}
 JH had the overall project leadership and introduced the main concepts together with MA. JN led the O data collection and the overall presentation of  the data in the manuscript. ST contributed to the preparation of the manuscript. All authors were involved in discussions and in preparing the manuscript. 
 
 \section*{Conflict of interest}
 The authors declare no conflict of interest.


\bibliography{JH-Cuprate.bib}
\printindex
\end{document}